\documentclass[sigconf]{acmart} 

\usepackage{graphicx}
\usepackage{caption}
\usepackage{subcaption}
\usepackage{natbib}
\usepackage{multirow}
\usepackage{tabularx}
\usepackage{makecell}

\AtBeginDocument{%
  \providecommand\BibTeX{{%
    \normalfont B\kern-0.5em{\scshape i\kern-0.25em b}\kern-0.8em\TeX}}}

\setcopyright{acmcopyright}
\copyrightyear{2024}
\acmYear{2024}
\acmDOI{XX.XXXX/XXXXXXX.XXXXXXX}

\acmConference[SBES'24]{Brazilian Symposium on Software Engineering}{September 30 -- October 04, 2024}{Curitiba, PR}


\setcopyright{none} 
\renewcommand\footnotetextcopyrightpermission[1]{} 

\begin{document}

\title[Revisiting Aristotle vs. Ringelmann]{Revisiting Aristotle vs. Ringelmann: The influence of biases on measuring productivity in Open Source software development}

\author{Christian Gut}
\email{christian.gut@usp.br}
\orcid{0009-0003-1627-1710}
\affiliation{%
  \institution{University of São Paulo}
  \city{São Paulo}
  \country{Brazil}
}

\author{Alfredo Goldman}
\email{gold@ime.usp.br}
\orcid{0000-0001-5746-4154}
\affiliation{%
  \institution{University of São Paulo}
  \city{São Paulo}
  \country{Brazil}
}

\renewcommand{\shortauthors}{Christian Gut and Alfredo Goldman}

\begin{abstract}
Aristotle vs. Ringelmann was a discussion between two distinct research teams from the ETH Zürich who argued whether the productivity of Open Source software projects scales sublinear or superlinear with regard to its team size. This discussion evolved around two publications, which apparently used similar techniques by sampling projects on GitHub and running regression analyses to answer the question about superlinearity. Despite the similarity in their research methods, one team around Ingo Scholtes reached the conclusion that projects scale sublinear, while the other team around Didier Sornette ascertained a superlinear relationship between team size and productivity. In subsequent publications, the two authors argue that the opposite conclusions may be attributed to differences in project populations, since 81.7\%~of Sornette's projects have less than 50~contributors. Scholtes, on the other hand, sampled specifically projects with more than 50~contributors.

This publication compares the research from both authors by replicating their findings, thus allowing for an evaluation of how much project sampling actually accounted for the differences between Scholtes' and Sornette's results. Thereby, the discovery was made that sampling bias only partially explains the discrepancies between the two authors. Further analysis led to the detection of instrumentation biases that drove the regression coefficients in opposite directions. These findings were then consolidated into a quantitative analysis, indicating that instrumentation biases contributed more to the differences between Scholtes' and Sornette's work than the selection bias suggested by both authors.  
\end{abstract}


\maketitle

\section{Introduction}
\label{sec:introduction}

With the ever-growing importance of software in modern society, the question of how to develop software as efficiently as possible has been relevant for decades. A subquestion on that topic, whether adding more developers to Open Source projects will generate economies or diseconomies of scale, had been investigated by two different research teams from the ETH Zürich. 

First, Sornette et al. deployed a regression analysis on the commit history of GitHub projects and identified a superlinear relationship between team size and team production~\cite{sornette_how_2014}. Their metric for production was the number of commits, or more precisely the number of edited files in each commit. This measurement enabled the calculation of a linear regression coefficient, denominated $\beta$, between the logarithm of the output of the team and the logarithm of the size of the team. The deployment of a 250-day period for each analysis and 5-day windows for the generation of data points allowed the authors to obtain multiple regression analyzes per project. The thereby identified superlinearities were associated with the heavy-tailed nature of the underlying distribution of contributions. Sornette et al. also calculated the mean values for each project, denominated as \textit{average~$\beta$}, and reported 104~out of 164~projects as being superlinear using this measure.

Almost 2~years later, Scholtes et al. detected diseconomies of scale in GitHub projects using an apparently similar regression analysis~\cite{scholtes_aristotle_2016}. Their approach to measuring production was more fine-grained by calculating the Levenshtein distance for each file edit. This editing distance was then used to calculate the team's productivity, using 7-day windows for the output and a 295-day windows with 7-day increments to determine the team size. Based on these data points, the researchers conducted several regression analyzes, considering Log-Log and Log-Lin models, and accounting for the influence of one-time contributors. The resulting regression coefficients were denominated as $\alpha_3$. For the case of the Log-Log regression without filtering for one-time contributors, Scholtes et al. discovered solely sublinear relationships across the 58~projects they examined.

The fact that both research teams reached opposite conclusions led to a follow-up publication that provided additional analytical insights~\cite{maillart_aristotle_2019}. An important argument of that study was that completely different sets of projects were examined, since 134~out of 164~projects analyzed by Sornette et al. had less than 50~contributors and Scholtes et al. deliberately sampled projects with more than 50~contributors. A study by Gote et al., based on Scholtes' original work, also suggested that the difference may be caused by the selection of projects~\cite{gote_big_2022}.

This work contributes to the Aristotle vs. Ringelmann discussion by addressing the following research question: 

\begin{quote} \textit{What factors explain the different conclusions drawn by Sornette et al. and Scholtes et al.?} \end{quote}

Since the two publications not only varied in their choice of projects, but also implemented comparable yet distinct regression techniques, the above research question can be split into these three more specific subquestions:

\begin{enumerate}
    \item \textit{How significant was project selection in creating the differences observed between Sornette et al. and Scholtes et al.?}
    \item \textit{What bias in Sornette's regression method could have favored the identification of a superlinear relationship between team size and productivity?}
    \item \textit{What bias in Scholtes' regression method could have favored the identification of a sublinear relationship between team size and productivity?}
\end{enumerate}

The first question was answered by reproducing both regression methods, so that the method of one author can be applied to the other authors' data set. Such an analysis gave quantitative evidence on how much project selection caused the opposite conclusions drawn by Sornette and Scholtes. A response to the second question led to a closer examination of Sornette's regression method, resulting in the discovery of a potential $p$-value filter that eliminated all regression coefficients close to zero. Similarly, the third question inspired an exploration into how the overall time frame selection affected Scholtes' regression method, revealing a strong influence of the first couple of days on the regression coefficient. 

The remainder of this paper is structured as follows. After a short summary of the related work on software development productivity, a comprehensive explanation of the project selection and regression methods used by Sornette and Scholtes is presented.  Subsequent sections show how the research methods were reproduced and how replication led to the conclusion that selection bias only partially accounts for the disparities between both authors. Next, quantitative evidence is provided demonstrating how instrumentation biases may have had a significant impact, and a conclusion summarizes how accounting for project selection and instrumentation biases approximates results. The final considerations provide a discussion about the lessons learned, limitations, and future work. 

\section{Related Work and State-of-the-Art}
\label{sec:related_work}

Arguably, the most foundational work on the productivity of software development teams is Brooks' book \textit{The Mythical Man Month}~\cite{brooks_mythical_1995}, from which the term \textit{Brooks Law} originated. This law states that adding people to a late software project will delay the project even further. Brooks also advocated for surgical teams that consist of a few highly specialized developers making most of the changes to the code, supported by a wide variety of team members who make the work of the few specialists as productive as possible. Another highly referenced classic work on software development productivity is the COCOMO II cost estimation model by Boehm et al.~\cite{boehm_cost_1995}. Although parts of this model might be outdated, the underlying factors like code reuse, the use of tools, or the team size, remain relevant. For the latter, Boehm identified diseconomies of scale.

Recent literature reviews confirm that the impact of team size on productivity in software development continues to be relevant. Chapetta and Travassos conducted a literature review to derive factors that influence developer productivity~\cite{chapetta_towards_2020}. At the beginning of their investigation, the authors examined 121~publications related to this topic. Then, they identified relevant factors and mapped their direction and intensity. Chapetta and Travassos also evaluated the belief in these factors by looking at the type and quality of the underlying study. These data points were then combined into a unified \textit{Gain in Belief} metric. The result of this exercise led to the discovery of 16~factors which influence software development productivity, one of which states a slightly adverse effect of team size on developer productivity. 

Another review of the literature on software development productivity comes from Duarte, who examined a total of 99~publications~\cite{duarte_software_2022}. He presented a synopsis of each article, providing valuable information on the potential underlying factors that affect software development productivity. These factors were related to size \& structure of the team, knowledge \& experience of the developers, development context, process \& tools, code structure, and social \& emotional factors. The study confirmed a significant negative impact of the size of the project on productivity, while team size exhibited a moderate negative influence.

Contemporary research also conducted quantitative analysis of the productivity of software development. Lavazza et al. is a publication based on the ISBG\footnote{International Software Benchmarking Group} data set~\cite{lavazza_empirical_2018}. This set consists of industry data that quantifies the output of the software development process in function points, which includes additional information, such as the programming language used, the industry involved, and whether a data point pertains to a new project or the maintenance of an existing one. The level of data quality in the ISBG data set varies widely, leading to the removal of more than 50\%~of the data points from the analysis. On the basis of the remaining data points, Lavazza et al. worked on two research questions: \textit{What factors influence the productivity of software development?} \textit{And what factors influence economies or diseconomies of scale?} In most cases, they were unable to identify economies or diseconomies of scale with statistical confidence. Exceptions were projects written in Java and Visual Basic, as well as certain types of enhancement projects, for which the authors detected economies of scale. Projects written in PL/I were the only exception where diseconomies of scale could be measured. 

A different analysis of software development productivity was conducted by Muríc et al., who specifically investigated projects on GitHub with fewer than 20~contributors~\cite{muric_collaboration_2019}. Within these limitations, the authors found that Open Source projects scale superlinear. However, this effect started to decrease once the team size exceeded 10~team member. An interesting aspect of their study is the notion of an effective team size $n$. This measurement is defined by $n = 2^H \text{, where } H = - \sum_{i=1}^N f_i \cdot log_2(f_i) \text{ and } f_i = w_i / W$, with $N$ being the number of team members, $w_i$ being the work done by the team member $i$, and $W$ being the sum of the work performed by the entire team. The effective team size reaches its maximum, $n = N$, if all members of the team contribute the same amount of work. Muríc et al. identified a negative correlation between effective team size and productivity, meaning that productivity increases if the work load is primarily handled by a few developers, who incorporate sporadic contributions from the rest of their team members. Interestingly, such a team structure has resemblance to the surgical teams described by Brooks.

Last but not least, there are two articles that can be directly linked to the Aristotle vs. Ringelmann discussion. First, there is a direct comparison of the two publications by Maillart and Sornette~\cite{maillart_aristotle_2019}. They argue that the opposite conclusions drawn by Sornette et al. and Scholtes et al. may be attributed to variations in the project population, because the superlinearity found in Sornette et al. holds for no more than 30~to 50~developers, and Scholtes et al. sampled specifically projects with more than 50~developers. Maillart and Sornette also emphasized that their work linked the phenomenon of superlinearity to the statistical distribution of contributions, which showed a heavy-tailed power-law behavior. This behavior can be explained by two possible generating mechanisms, one based on cascading interactions and the other involving large deviations of the underlying process. They further asserted that the link between these generating mechanisms and superlinear production was validated in their work by looking at bursts of activity per contributor. Since the frequency of these bursts of activity decreases with larger projects, and given that Scholtes' data set mainly consists of large projects, the authors suggested that these factors could explain the divergent conclusions drawn by the two research groups. In the end, Maillart and Sornette highlighted that the primary focus of their work was different from Scholtes' research, since they examined bursts of production while Scholtes considered averages. Nevertheless, it should be mentioned that, despite focusing on bursts of production, even the averages reported by Sornette et al. showed superlinearity, which justifies the investigations conducted in this study.

The other publication related to the Aristotle vs. Ringelmann discussion was conducted by Gote et al., which refined Scholtes' initial work by introducing team size stratification~\cite{gote_big_2022}. This work also provided new insights into the correlation of output metrics, economies of scale, and the influence of foreign code edits on productivity. Initially, the authors identified a high correlation between multiple output metrics, such as the number of commits, the Levenshtein distance, changes in LOC, or changes in cyclomatic complexity. This implies that the different output metrics may only have a marginal influence on the results when analyzing the productivity of software development teams. Gote's approach to stratify their analysis by team size led to the identification of slight economies of scale for teams with less than 20~contributors. For larger projects, diseconomies of scale prevailed. Based on this finding, they argued that any study favoring small projects in its sampling may mistakenly infer the presence of overall economies of scale. Finally, they identified the ratio of foreign code edits, which is the percentage of modifications to source code that has not been written by the author itself, as a factor that negatively correlates with productivity. This conclusion not only seems logical, but could also provides a rationale for the occurrence of diseconomies of scale, as having a larger team working on a project may lead to an increased likelihood of foreign code edits.

This study adds to the latest research by demonstrating that selection bias alone is not a sufficient explanation for the Aristotle vs. Ringelmann debate. It further illustrates how instrumentation biases appear to be a relevant factor. Identifying these instrumentation biases not only contributes to the specific debate between Sornette and Scholtes but also hints at general factors that must be considered when examining the productivity of Open Source software development.

\section{Differences in Sornette's and Scholtes' work}
\label{sec:difference_sornette_scholtes}

In general terms, Sornette et al. and Scholtes et al. used the same regression approach to determine superlinearity: 

\begin{enumerate}
    \item Select a set of projects
    \item Divide an overall time frame for each project into smaller time windows
    \item For each time window
    \begin{enumerate}
        \item Determine the team size
        \item Determine the output produced by the team
    \end{enumerate}
    \item Plot output and team size on a graph
    \item Run a linear regression analysis
\end{enumerate}

\begin{figure*}
\includegraphics[width=\linewidth]{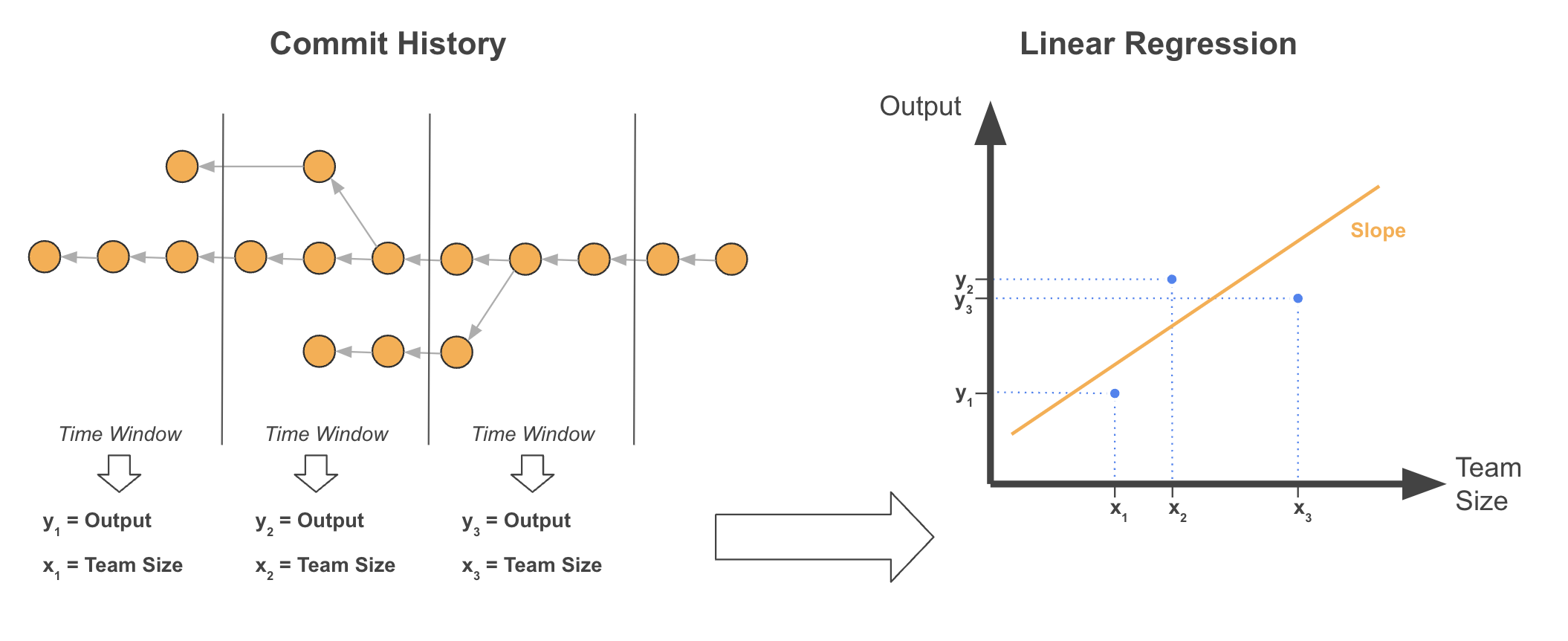}
\caption{General regression approach used by both authors}
\Description{Illustration how the data from the git commit history is being transformed into a regression analysis.}
\label{fig:regression_method_general}
\end{figure*}

Figure~\ref{fig:regression_method_general} is a visualization of this general approach. However, these approaches differ significantly in terms of their implementation specifics.

\subsection{Project selection}

Both articles seemed to have made a similar choice in terms of data collection: Mine all commits from a set of GitHub projects starting with the very first commit and ending with some cut-off date in the middle of the 2010's.

Nevertheless, a closer examination reveals some fundamental differences in the characteristics of the selected projects. Sornette et al. looked at 164~projects while Scholtes et al. examined only 58~projects. Intriguingly, only 3~projects were analyzed by both authors: Rails, Django, and jQuery. This represents less than 1.4\%~of all 219~projects. Furthermore, Sornette et al. opted for a random sampling approach, while Scholtes et al. made the deliberate choice to select only projects with more than 50~contributors.

\subsection{Regression method}

\begin{figure*}    

\begin{subfigure}[b]{\linewidth}
\includegraphics[width=\linewidth]{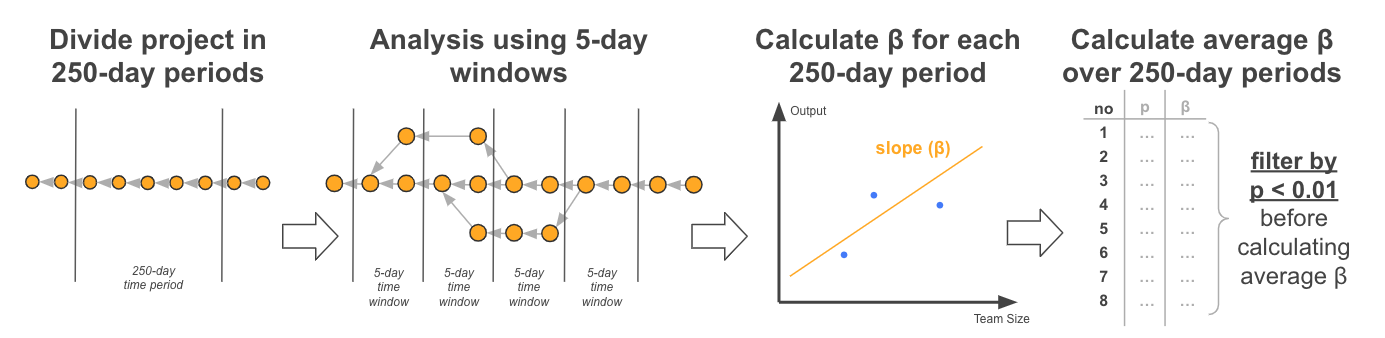}
\caption{Illustration of Sornette's regression method}
\Description{The four steps in Sornette's regression method}
\label{fig:regression_method_details_sornette}
\end{subfigure}

\begin{subfigure}[b]{\linewidth}
\includegraphics[width=\linewidth]{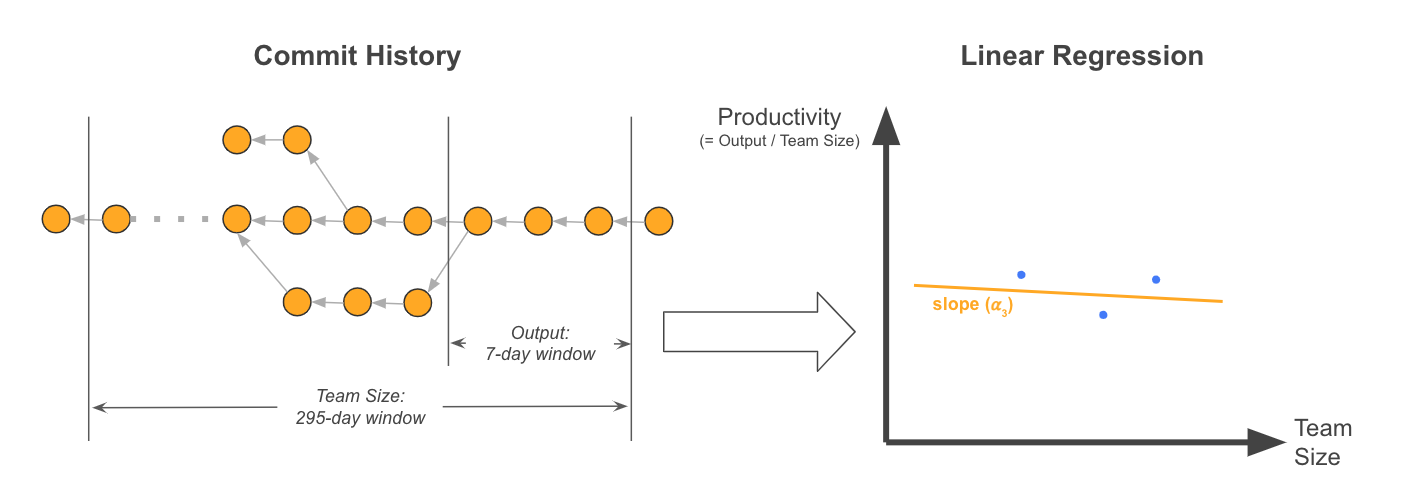}
\caption{Illustration of Scholtes' regression method}
\Description{The two step approach used by Scholets' regression method}
\label{fig:regression_method_details_scholtes}
\end{subfigure}

\caption{Illustration regression method details}
\label{fig:regression_method_details}
\end{figure*}

The authors differed not only in the selection of projects, but also in the details of their regression analysis. These differences are illustrated in Figure~\ref{fig:regression_method_details} and will be explained in more detail below.

\subsubsection{Definition of time windows}

Sornette et al. divided the project first into 250-day periods and then subdivided each 250-day period into 5-day windows. The team output and team size were determined for each 5-day window. 

Scholtes et al. divided all data for a project into 7-day windows. These 7-day windows were used to calculate the team output. Remarkably, the team size itself had been determined by using 295-day windows which were incremented by 7-day intervals that match the end date of the team output window for each iteration. This implies that the team size windows overlapped, so that a commit was included multiple times when determining team sizes, but was only counted once when calculating team output.

\subsubsection{Measurement of output}

Sornette et al. reported commit count as an output measure, but an evaluation of Sornette's Numpy files on PLOS suggested that the authors actually worked with the number of modified files in every commit. This assumption could be confirmed during a subsequent email exchange with Thomas Maillart, one of the coauthors.

Scholtes et al. used the Levenshtein editing distance to determine the output of the team. This distance measure was applied over all non-binary files by comparing the file's content before the commit against the file's content after the commit. Depending on the particular case, it can make a difference if the distance is calculated on a line-by-line, block-by-block, or file-by-file basis. Since inferring the exact method was not possible, an email exchange with Ingo Scholtes was initiated, who provided assistance by recommending the use of the line-by-line calculation method for the replication. 

\subsubsection{Measurement of team size}

To determine the team size, each team member must be uniquely identified. For this purpose, a Git commit provides four possible values: \textit{Author Email}, \textit{Author Name}, \textit{Committer Email}, or \textit{Committer Name}. In the majority of instances, the author and committer fields are the same. However, if modifications are made to a commit (e.g. via the \textit{git rebase} command), the committer fields may differ from the author fields. This study used \textit{Author Email} as the unique identifier of a team member. Scholtes et al. clearly stated the same choice in their publication, and an examination of the data on PLOS indicated that Sornette et al. also opted for this approach.

\subsubsection{Regression analysis}

Both authors determined if a project scales superlinearly by running a linear regression. However, variations in the setup of the regression analyzes have been observed:

\begin{itemize}

    \item Sornette et al. defined the final metric for a project as an aggregation of multiple regression coefficients, called \textit{average~$\beta$}. Scholtes et al. calculated the coefficient of a simple linear regression, which they denominated $\alpha_3$.
    
    \item Sornette et al. plotted the team's output against the team size while Scholtes et al. compared the team's productivity with the team size. Since Scholtes' productivity metric divides the team's output by team size, this difference is negligible in practical terms. It only means that Scholtes' method identifies superlinear projects for a slope $> 0$, while the same applies to Sornette`s method for a slope $> 1$.
    
    \item Sornette et al. divided each project into non-overlapping 250-day periods before running a regression analysis on each of them, while Scholtes et al. used all available data points for one simple regression analysis per project.
    
    \item Scholtes et al. provided an additional analysis for Log-Lin relationships in which the logarithm of the productivity measure was plotted against the plain unmodified team size.
    
    \item Scholtes et al. also controlled for one-time contributors by providing an analysis that excluded contributions from anyone who submitted only one commit.
    
    \item Sornette et al. took statistical significance into account by calculating the \textit{average~$\beta$} only over those 250-day periods where the $p$-value was $p < 0.01$. 
    
\end{itemize}

It should also be noted that the $p < 0.01$ filter identified in Sornette's method had only been detected while attempting to replicate the results. Without using this filter, reproduction of the \textit{average~$\beta$} values was not possible. Even minor alterations, such as a $p < 0.05$ filter, resulted in significantly higher $p$-values for the statistical tests, corroborating with the hypothesis that the $p < 0.01$ filter was actually part of the original study. This filter also implies that \textit{average~$\beta$} cannot be determined if the filter causes an exclusion of all $\beta$~values. 

\subsection{Summary}

The two authors made quite distinct design choices, which are summarized in Table~\ref{tab:comparison_scholtes_sornette}. Sornette et al. opted for a random sampling of projects and a simpler approach in measuring the team output, but adopted a more elaborated method in deriving the time windows for measuring the regression coefficients. Scholtes et al. chose a more straightforward method to calculate the regression coefficients, but selected projects more deliberately and studied more variants of their regression method by looking into Log-Lin regressions and controlling for one-time contributors. They also provided a more fine-grained measurement of the team's output. However, the influence of this fine-grained measure may not be as relevant due to the high degree of correlation between different output metrics identified by Gote et al.~\cite{gote_big_2022}.

\begin{table}

\caption{Comparison of projects analyzed and regression method used by Sornette et al. and Scholtes et al.}

\begin{tabular}{ lll  }
 \toprule

 & \thead[l]{Sornette et al.} & \thead[l]{Scholtes et al.} \\
 \midrule

\multirow{2}{*}{\thead[l]{Projects}} & \thead[l]{164~projects (3~overlapping)} & \thead[l]{58~projects (3~overlapping)} \\
 & \thead[l]{5-3,074 contributors} & \thead[l]{55-4,107 contributors} \\
 \midrule

\multirow{9}{*}{\thead[l]{Regression\\ method}} & \thead[l]{Output: 5-day windows} & \thead[l]{Output: 7-day windows} \\
 & \thead[l]{Team size: 5-day windows} & \thead[l]{Team size: 295-day windows} \\
 \cmidrule(lr){2-3}
 & \thead[l]{No. of file edits} & \thead[l]{Levenshtein distance}  \\
 \cmidrule(lr){2-3}
 & \thead[l]{Over 250-day periods} & \thead[l]{Over whole project} \\
 & \thead[l]{Average of regression slopes} & \thead[l]{One regression slope only} \\
 & \thead[l]{Filtered by $p < 0.01$} & \thead[l]{No filtering}  \\
 & \thead[l]{Log-Log regression only} & \thead[l]{Log-Log and Log-Lin} \\ 
 & \thead[l]{No control for any variable} & \thead[l]{Control for\\one-time contributors} \\ 
\bottomrule
\end{tabular}

\label{tab:comparison_scholtes_sornette}
\end{table}

\section{Replication of Sornette's and Scholtes' work}
\label{sec:reproduction}

To ensure that any extrapolation of the original study is free from errors, it is essential to confirm that the original findings of Sornette et al. and Scholtes et al. can be replicated with statistical confidence. Shull et al. refer to this practice as performing a dependent replication, with the aim of closely reproducing the original findings, and recommend that it should come before any conceptual replications, which involve altering parameters to gain new insights~\cite{shull_role_2008}. Therefore, this section provides an in-depth review of the replication process before exploring possible selection and instrumentation biases.

\subsection{Technical implementation}

The technical implementation of this study is based on a set of Python scripts, using the PyDriller library~\cite{spadini_pydriller_2018}. These scripts ran on Google's Colab notebooks, using BigQuery as a storage engine.  This choice had been made for reasons of convenience, with the objective of reducing the amount of time required to configure systems and maintain an IT~infrastructure.  

Another approach would have been the use of \textit{git2net}, a tool developed by Scholtes' team~\cite{gote_git2net_2019}. This tool provides additional network analysis capabilities that are not needed for this research. Since these additional capabilities imply significantly longer execution times, the creation of a simpler and more performant mining script appeared to be the better option for data collection.

The creation of own tools comes with the risk of introducing errors. For example, Levenshtein distance calculations might have been performed differently than in the original tools. To mitigate these risks, the output of the scripts had been compared with the data provided by \textit{git2net}, confirming that both tools compute exactly the same distances for all relevant, non-merge commits. 

\begin{figure}
\includegraphics[width=\linewidth]{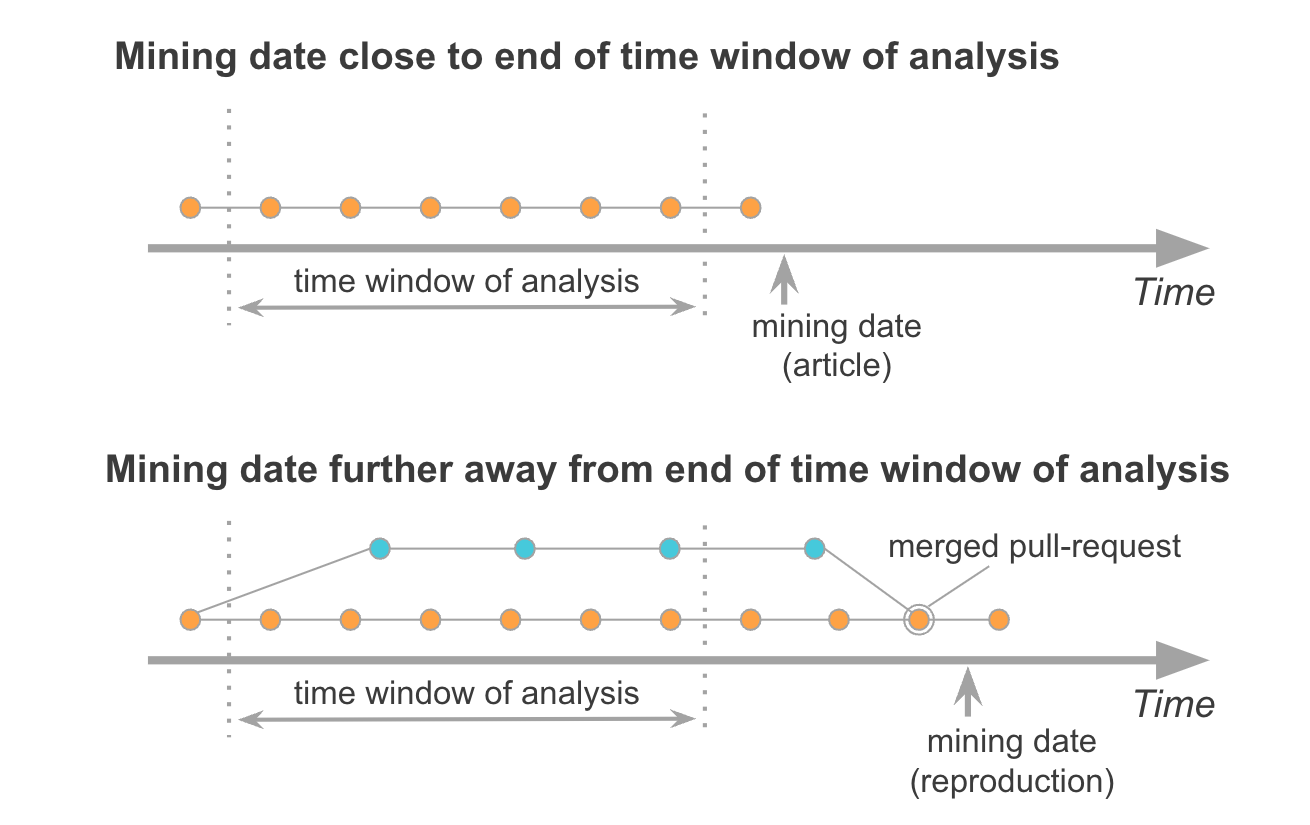}
\Description{Illustration how pull requests which are merged after the mining date lead to the exclusion of relevant commits.}
\caption{Potential exclusion of relevant commits if time window end date and mining date lie too close together}
\label{fig:illustration_influence_of_pull_requests}
\end{figure}

GitHub itself bears some pitfalls when it comes to the reproducibility of results. In particular, the \textit{git rebase} command allows for a retroactive modification of commits. These modifications can increase or decrease the number of commits, which is a known problem with little remediation when it comes to the reproduction of studies. Furthermore, pull requests may cause distortions if the mining date is close to the end date of the time window of analysis, as demonstrated in Figure~\ref{fig:illustration_influence_of_pull_requests}. In these cases, merging a pull request could add relevant commits to a repository that were not previously visible to the researcher. Finally, some very special cases of manual amendments might put the commit dates out-of-order, thus tripping off date filters, which assume all commits to be ordered by date.

The reproduction of results requires the exact GitHub URLs, which were not provided by neither author. So, a manual mapping of project names to GitHub URLs had to be performed. For Sornette et al., it was possible to identify the URL for 159~out of 164~projects. In the case of Scholtes et al. all 58~projects could be mapped to an URL. 

Reproducing results also necessitates identifying the correct time windows of analysis for each GitHub project. For Sornette et al. the dates could be inferred by extracting the number of days of analysis from the Numpy data files published on PLOS, and by working with the assumption that the first commit on GitHub equals the start date of the analysis. In Scholtes' case the exact start and end dates of the analysis were part of the article's appendix. 

\subsection{Statistical tests}

All of the factors in the above section are reasons why the replication of precise figures can be challenging, as was the case in this study. Therefore, a pairwise comparison was carried out to assess whether the measured and reported values for each project were drawn from the same distribution. Thereby, the Wilcoxon test was performed since the Shapiro-Wilk test indicated that none of the regression coefficients adhered to a normal distribution~\cite{wilcoxon_individual_1945}\cite{shapiro_analysis_1965}. In both cases, the SciPy Python package had been leveraged~\cite{virtanen_scipy_2020}.

Both authors not only reported the regression coefficients in their \textit{Article}, but also provided \textit{Commit Data} via PLOS or Zenodo. These data sets in addition to the data mined on \textit{GitHub} allow for the following pairwise comparisons:

\begin{enumerate}
    \item \textit{Article vs. GitHub}: The ultimate test if the results can be reproduced.
    \item \textit{Article vs. Commit Data}: A check if the regression calculations are being performed correctly.
    \item \textit{Commit Data vs. GitHub}: An additional check of consistency.    
\end{enumerate}

All these comparisons used Scholtes' Log-Log regression model without a filter of one-time contributors, since this variant is closest in resemblance to Sornette's method. 

\begin{figure}

\begin{subfigure}[b]{\linewidth}
\includegraphics[width=\linewidth]{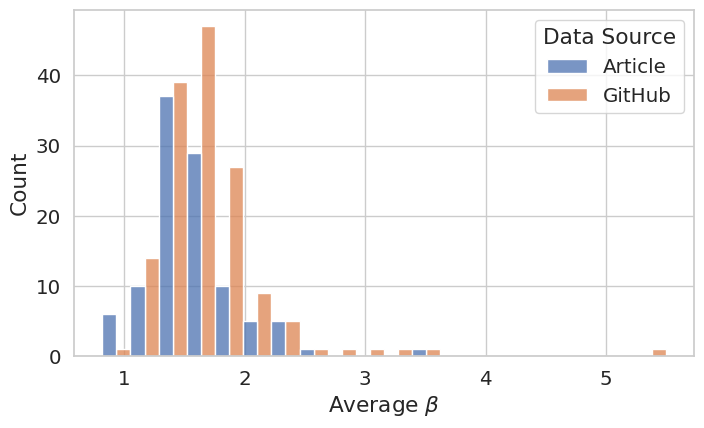}
\Description{Histogram comparing the average beta values of the original publication with the reproduced values.}
\caption{Results for Sornette's \textit{average~$\beta$}}
\label{fig:histogram_github_article_sornette}
\end{subfigure}
\hfill
\begin{subfigure}[b]{\linewidth}
\includegraphics[width=\linewidth]{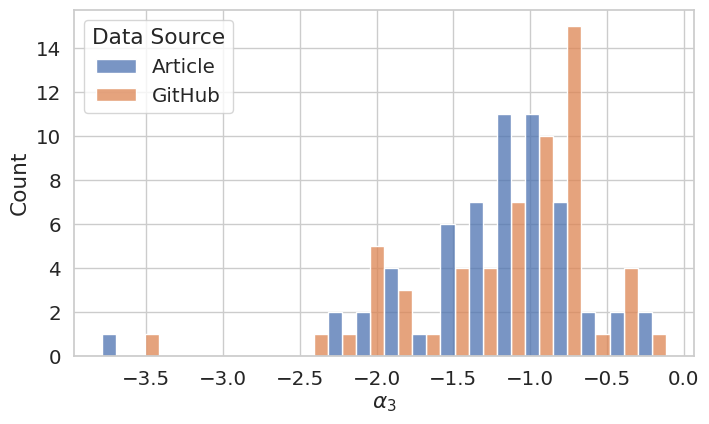}
\Description{Histogram comparing the $\alpha_3$ values of the original publication with the reproduced values.}
\caption{Results for Scholtes' $\alpha_3$}
\label{fig:histogram_github_article_scholtes}
\end{subfigure}

\caption{Comparison between regression coefficients as reported in the article and the measured values using data from GitHub}
\label{fig:histogram_github_article}

\end{figure}

\begin{figure}

\begin{subfigure}[b]{0.48 \linewidth}
\includegraphics[width=\linewidth]{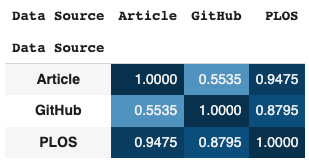}
\Description{Table of the results from the Wilcoxon statistical tests for Sornette et al.}
\caption{Results for Sornette et al.}
\label{fig:results_wilcoxon_sornette}
\end{subfigure}
\hfill
\begin{subfigure}[b]{0.48 \linewidth}
\includegraphics[width=\linewidth]{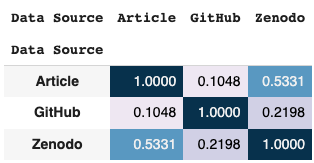}
\Description{Table of the results from the Wilcoxon statistical tests for Sornette et al.}
\caption{Results for Scholtes et al.}
\label{fig:results_wilcoxon_scholtes}
\end{subfigure}

\caption{Results of the Wilcoxon tests}
\label{fig:results_wilcoxon}

\end{figure}

The histogram in Figure~\ref{fig:histogram_github_article} compares the data reported in the \textit{Article} and the reproduced results using data from \textit{GitHub}. The apparent similarity of the distributions could be confirmed by Wilcoxon tests, whose values are depicted in Figure~\ref{fig:results_wilcoxon}. In all cases, the null hypothesis that the data came from different distributions could not be rejected, implying that the results were similar and that the replicated data could be used for further analysis.

\section{Selection Bias}
\label{sec:select_bias}

Subsequent research indicated that the differing conclusions reached by Sornette et al. and Scholtes et al. can be attributed to the selection of projects~\cite{maillart_aristotle_2019}\cite{gote_big_2022}. This section is an attempt to effectively quantify the impact of project selection by calculating Sornette's \textit{average~$\beta$} and Scholtes' $\alpha_3$ for all projects. If selection bias was significant, many of the projects chosen by Sornette et al. would produce values of $\alpha_3 > 0$, while the projects selected by Scholtes et al. would generate values of \textit{average $\beta < 1$}.  

\begin{figure}

\begin{subfigure}[b]{\linewidth}
\includegraphics[width=\linewidth]{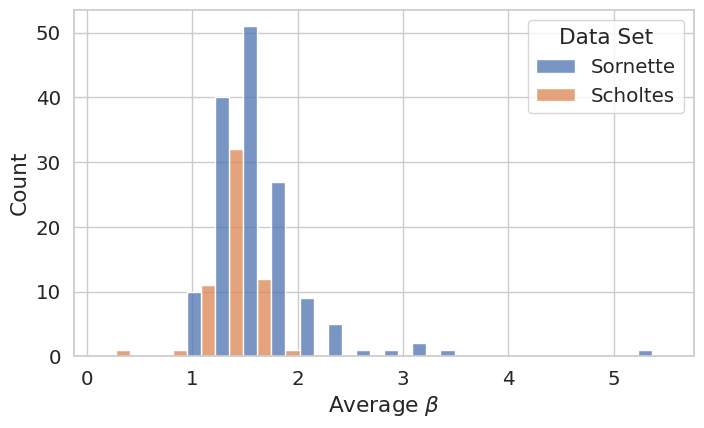}
\Description{Histogram of the average beta values using Sornette's and Scholtes' data sets.}
\caption{Comparison for Sornette's average~$\beta$ values}
\label{fig:comparison_projects_sornette}
\end{subfigure}
\hfill
\begin{subfigure}[b]{\linewidth}
\includegraphics[width=\linewidth]{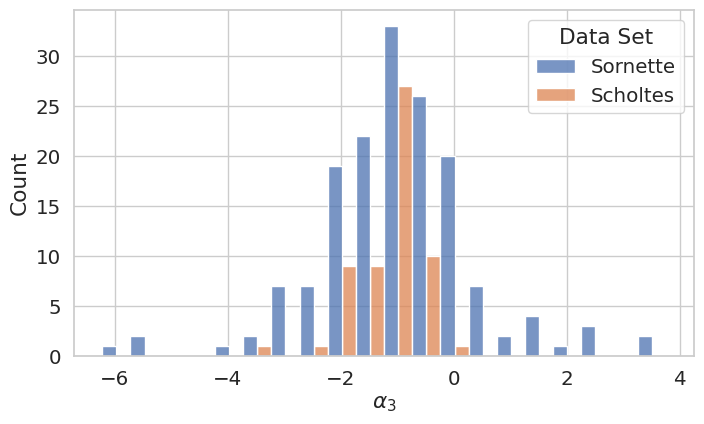}
\Description{Histogram of the $\alpha_3$ values using Sornette's and Scholtes' data sets.}
\caption{Comparison for Scholtes' $\alpha_3$ values}
\label{fig:comparison_projects_scholtes}
\end{subfigure}

\caption{Cross-comparison of data sets and regression methods}
\label{fig:comparison_projects}

\end{figure}

The result of this exercise is shown in Figure~\ref{fig:comparison_projects}. For both regression methods, the distributions generated by each data set seem to differ, which could be confirmed by the $p$-values of two-sample Kolmogorov-Smirnov tests using the SciPy package~\cite{hodges_significance_1958}\cite{virtanen_scipy_2020}: For \textit{average~$\beta$} a $p$-value of $p = 0.00000000099$ was obtained and for $\alpha_3$ a value of $p = 0.001383$ was measured, indicating that the results from both data sets differed significantly for either method.

However, the information presented in Figure~\ref{fig:comparison_projects} and Table~\ref{tab:comparison_selection_bias} also suggests that the selection of projects only partially accounts for the differences between Sornette and Scholtes. For most of Scholtes' projects, a superlinear relationship had been detected, even when Sornette's method was applied. Likewise, Scholtes' method measured sublinear relationships for the majority of projects in Sornette's data set. Both facts clearly indicate that selection bias alone is not a sufficient explanation for the differences between Sornette et al. and Scholtes et al., necessitating further investigation of biases within the regression methods.

\begin{table}
\caption{Number of projects with sub- and superlinear relationship between team size and productivity}
\label{tab:comparison_selection_bias}
\begin{tabular}{ llccc  }
 \toprule
\thead[l]{Regression\\Method} & \thead[l]{Data Set} & \thead[c]{Sublinear\\Projects} & \thead[c]{Superlinear\\Projects}  & \thead[c]{Projects\\Total} \\
 \midrule
\multirow{2}{*}{\thead[l]{Sornette}} & \thead[l]{Sornette} & 1 & 147 & 148 \\
 \cmidrule(){2-5}
    & \thead[l]{Scholtes} & 3 & 55 & 58 \\
\midrule
\multirow{2}{*}{\thead[l]{Scholtes}} & \thead[l]{Sornette} & 130 & 29 & 159 \\
 \cmidrule(){2-5}
    & \thead[l]{Scholtes} & 58 & 0 & 58 \\
\bottomrule
\end{tabular}
\end{table}

\section{Instrumentation Biases}
\label{sec:instrumentation_biases}

The preceding section has already shown that project selection only partially accounts for the discrepancies between Sornette and Scholtes. This section eludes to potential biases found in either regression method, thus providing an additional explanation for the differences between Sornette's and Scholtes' findings.

\subsection{Sornette's p-value filter}

Section~\ref{sec:difference_sornette_scholtes} mentions that reproduction efforts for the work of Sornette et al. were only successful after applying a $p$-value filter with $p < 0.01$ when calculating the \textit{average~$\beta$} values for a project. However, these $p$-values tend to get high for slopes close to zero, because they represent results \textit{"for a hypothesis test whose null hypothesis is that the slope is zero, using Wald Test with t-distribution of the test statistic"}~\cite{the_scipy_community_linregress_2008}. Hence, the filter can introduce bias by excluding small values of $\beta$ from the computation of the \textit{average~$\beta$}.

\begin{figure}    
\includegraphics[width=\linewidth]{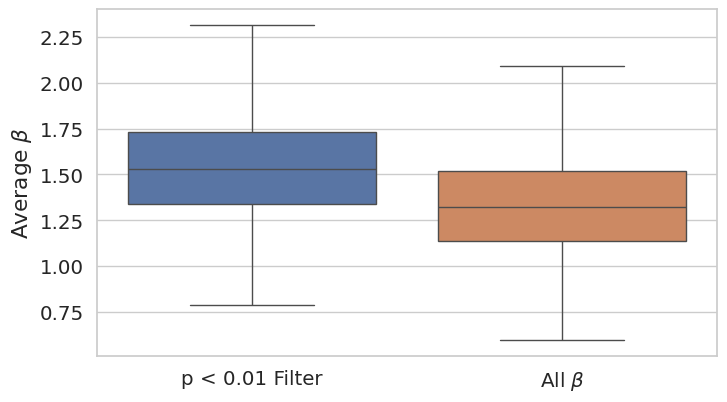}
\Description{Box-Whisker plot comparing the distribution of average beta values.}
\caption{Distribution of \emph{average~$\beta$}, comparing values using $p < 0.01$ to values which use all $\beta$}
\label{fig:average_beta_with_and_without_p_filter}
\end{figure}

Figure~\ref{fig:average_beta_with_and_without_p_filter} compares the distribution of \textit{average~$\beta$} values with a $p < 0.01$ filter to values without any filter. It can be clearly seen that the removal of the filter causes a reduction of \textit{average~$\beta$}, which was 14.82\%~on average. A Wilcoxon test with $p = 0.0000000$ also confirms that the reduction was of statistical significance. 

The effect of the filter also influences the sub- and superlinearity, as shown in Table~\ref{tab:superlinearity_p_value_filter}. For Sornette's data set the number of projects with diseconomies of scale increases from 1~to~14, and for Scholtes' data set the increase is from 3~to 16~projects. These numbers indicate that the $p$-value filter seemed to have a real influence on Sornette's conclusions.

\begin{table}
\caption{Effect of $p$-value filter on number of projects with sublinear relationship between team size and productivity}
\label{tab:superlinearity_p_value_filter}
\begin{tabular}{ llccc  }
 \toprule
\thead[l]{Regression\\Method} & \thead[l]{Data Set} & \thead[c]{Sublinear\\Projects} & \thead[c]{Superlinear\\Projects}  & \thead[c]{Projects\\Total} \\
 \midrule
\multirow{2}{*}{\thead[l]{Sornette\\($p < 0.01$)}} & \thead[l]{Sornette} & 1 & 147 & 148 \\
 \cmidrule(){2-5}
    & \thead[l]{Scholtes} & 3 & 55 & 58 \\
\midrule
\multirow{2}{*}{\thead[l]{Sornette\\(all $\beta$)}} & \thead[l]{Sornette} & 14 & 134 &  148 \\
 \cmidrule(){2-5}
    & \thead[l]{Scholtes} & 16 & 42 & 58 \\
\bottomrule
\end{tabular}
\end{table}

\subsection{Scholtes' time window selection}

\begin{figure}
\centering
\includegraphics[width=\linewidth]{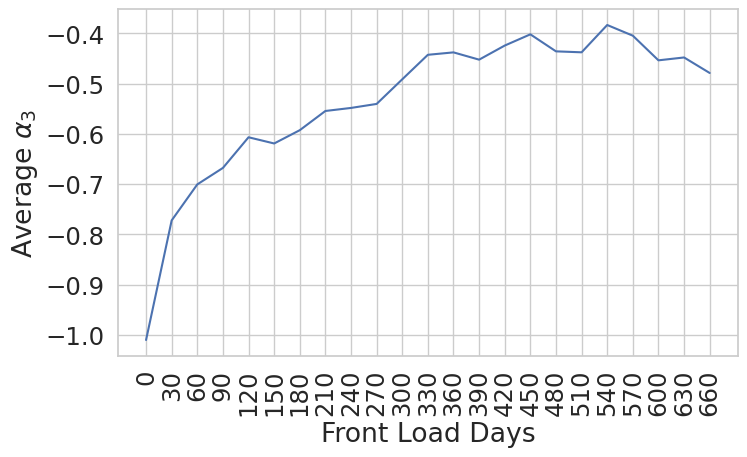}
\Description{Line graph showing the average $\alpha_3$ as a function of Front Load Days.}
\caption{The influence of \textit{Front Load Days} on the average of $\alpha_3$ over all projects}
\label{fig:alpha3_as_function_of_front_load_days}
\end{figure}

The regression analysis performed by Scholtes et al. took into account all contributions, from the very first commit until a certain cut-off date in 2014. An exploratory analysis indicated that removing the first few days from a project significantly influences the results of Scholtes' method, leading to the suspicion that the choice of time window may bear an instrumentation bias. A more detailed investigation led to Figure~\ref{fig:alpha3_as_function_of_front_load_days}, showing the effect on $\alpha_3$ if the first days of a project, denoted as \textit{Front Load Days}, are removed from the regression analysis. The steady increase until a peak at 540~days is evidence that Scholtes' method may indeed be affected by instrumentation bias.

\begin{figure}
\centering
\includegraphics[width=\linewidth]{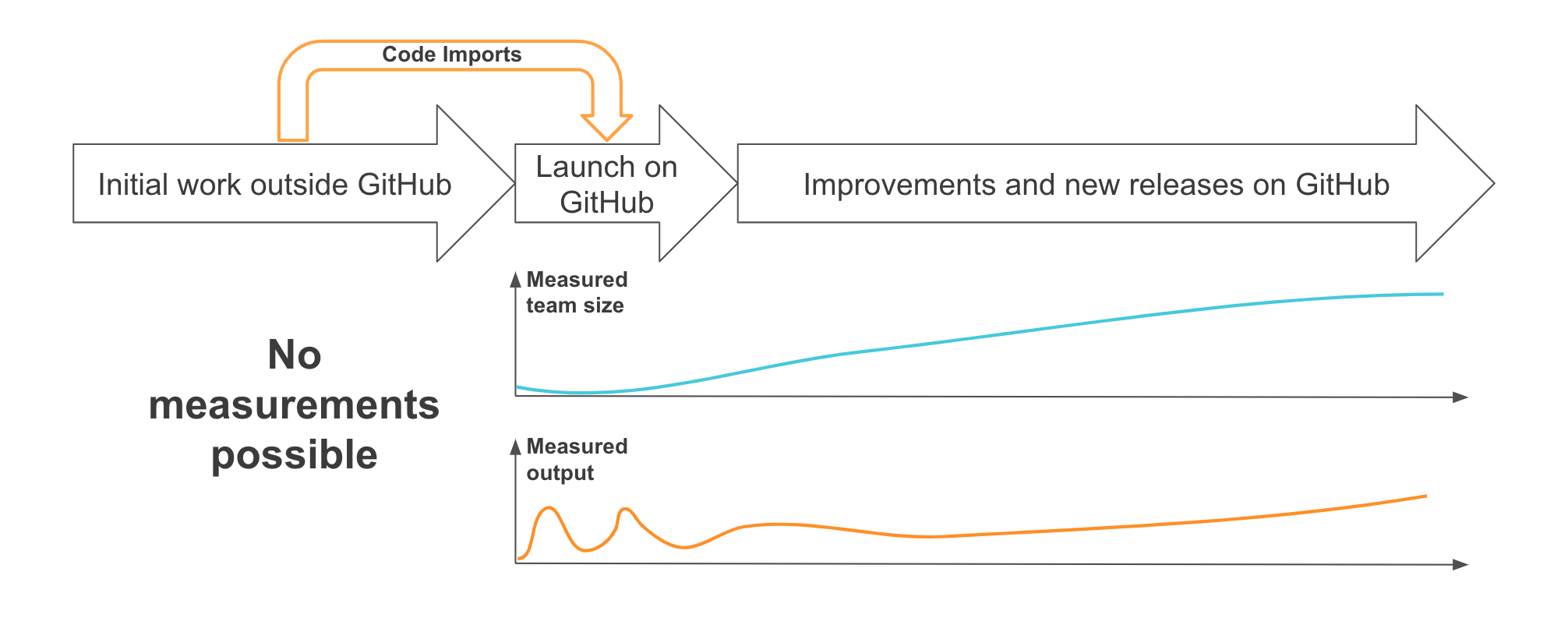}
\Description{Illustration how code imports of work before a launch on GitHub may explain the effect of the Front Load Days.}
\caption{Illustration of \textit{Front Load Days} effect: Code imports during the launch phase on GitHub may distort results}
\label{fig:illustration_effect_of_front_load_days}
\end{figure}

One possible explanation for this bias could be code imports during the first days of publication on GitHub. A developer or a team of developers may already have engaged in significant, non-measurable work before publishing the first commit. This work remains invisible until its inclusion via code imports after the project's launch. Since the early stages of a project tend to involve fewer developers, imports potentially generate artificial spikes of team output, which correlate with small team sizes, and thus lead to a negative bias for the regression analysis. Figure~\ref{fig:illustration_effect_of_front_load_days} illustrates this effect.

\begin{figure}
\centering
\includegraphics[width=\linewidth]{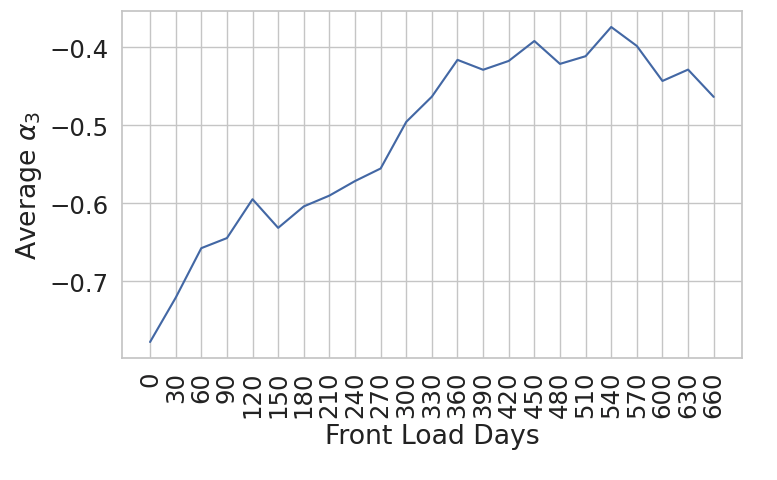}
\Description{Line graph showing the average $\alpha_3$ as a function of Front Load Days, when excluding Levenshtein distance outliers.}
\caption{The influence of \textit{Front Load Days} on the average of $\alpha_3$ over all projects, when excluding the top and bottom 2.5\%~of contributions based on Levenhstein distance}
\label{fig:alpha3_as_function_of_front_load_days_removing_outliers}
\end{figure}

The research of Gote et al. also addresses the impact of code imports by excluding the top and bottom 2.5\%~of contributions, in terms of Levenshtein distance~\cite{gote_big_2022}. However, the effect of \textit{Front Load Days} prevails even when filtering by these outliers, as shown in
Figure~\ref{fig:alpha3_as_function_of_front_load_days_removing_outliers}. A potential reason could be that code imports can happen in small chunks (i.e. module-by-module) and therefore would not entirely be filtered out by accounting for outliers. 

\begin{figure}
\centering
\includegraphics[width=\linewidth]{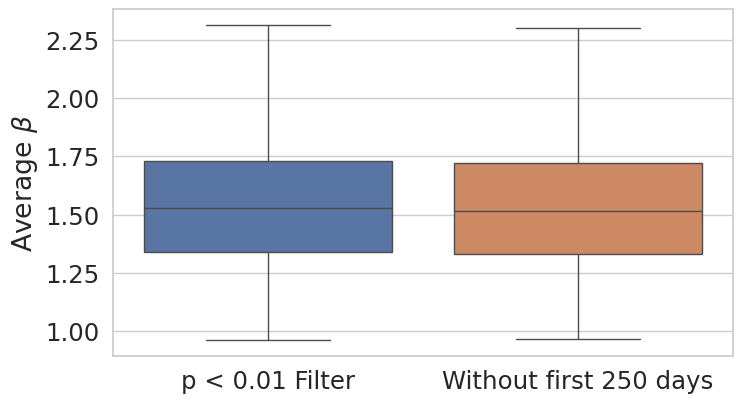}
\Description{Box-Whisker plot comparing the distributions of average beta.}
\caption{Comparison of original \textit{average~$\beta$} distribution with the values where the first 250-days were removed}
\label{fig:average_beta_without_first_250_days}
\end{figure}

The influence of \textit{Front Load Days} is negligible in the case of Sornette's regression method because the use of 250-day periods inhibits propagation of that bias throughout the entire regression analysis. A comparison of the \textit{average~$\beta$} values where the first 250-day period had been removed with the values of the original method, as shown in Figure~\ref{fig:average_beta_without_first_250_days}, demonstrates that the exclusion of the first couple of days had little influence on the overall outcomes. A Wilcoxon test with a $p$-value of $p = 0.9841$ supports this finding.

Eliminating \textit{Front Load Days} comes with the drawback of having to exclude certain projects from the analysis due to insufficient data points for the calculation of the regression coefficient. Consequently, the decision was made to choose 330~\textit{Front Load Days} to examine the impact on superlinearity. This choice eliminated less than 5\%~of all projects from the analysis. At the same time, 330~days seems to be a point where the average $\alpha_3$ appears to stabilize. For 330~\textit{Front Load Days} the number of projects with a superlinear relationship increases from 29~to 65~projects in the case of Sornette's data set, and from 0~to 11~projects in the case of Scholtes' data set, as shown in Table~\ref{tab:superlinearity_front_load_days}. This increase suggests that the choice of \textit{Front Load Days} has a relevant impact on the results of Scholtes' regression method.

\begin{table}
\caption{Effect of \textit{Front Load Days} on number of projects with sublinear relationship between team size and productivity}
\label{tab:superlinearity_front_load_days}
\begin{tabular}{ llccc  }
 \toprule
\thead[l]{Regression\\Method} & \thead[l]{Data Set} & \thead[c]{Sublinear\\Projects} & \thead[c]{Superlinear\\Projects}  & \thead[c]{Projects\\Total} \\
 \midrule
\multirow{2}{*}{\thead[l]{Scholtes\\(0~days)}} & \thead[l]{Sornette} & 130 & 29 & 159 \\
 \cmidrule(){2-5}
    & \thead[l]{Scholtes} & 58 & 0 & 58 \\
\midrule
\multirow{2}{*}{\thead[l]{Scholtes\\(330~days)}} & \thead[l]{Sornette} & 85 & 65 &  150 \\
 \cmidrule(){2-5}
    & \thead[l]{Scholtes} & 46 & 11 & 57 \\
\bottomrule
\end{tabular}
\end{table}

\section{Conclusion}
\label{sec:conclusion}

The discussion of Aristotle vs. Ringelmann evolved around the question whether the productivity of Open Source projects scales sublinear or superlinear with regard to its team size. The conflicting findings reported by Sornette et al. and Scholtes et al. prompted both research groups to publish follow-up studies, suggesting project selection as a possible explanation. The answer to the first research question of this study -- \textit{"How significant was project selection in creating the differences observed between Sornette et al. and Scholtes et al.?"} -- probes into the significance of the suggested selection bias by replicating both regression methods on both data sets. This analysis revealed that 130~of 159~projects from Sornette's data set scaled sublinearly using Scholtes' method. Likewise, 55~of 58~projects from Scholtes' data set showed superlinear scaling if subjected to Sornette's method. Thus, indicating that selection bias had only a minor influence.

This finding led to the second research question -- \textit{"What bias in Sornette's regression method could have favored the identification of a superlinear relationship between team size and productivity?"} -- which identified a $p$-value filter as a source of bias, whose removal increased the total number of projects that scale sublinearly from 4~to~30. Similarly, an inquiry into the third research question -- \textit{"What bias in Scholtes' regression method could have favored the identification of a sublinear relationship between team size and productivity?"} -- led to the discovery of \textit{Front Load Days}, whose removal increased the total number of projects that scale superlinearly from 29~to~76.

\begin{figure}
\centering
\includegraphics[width=\linewidth]{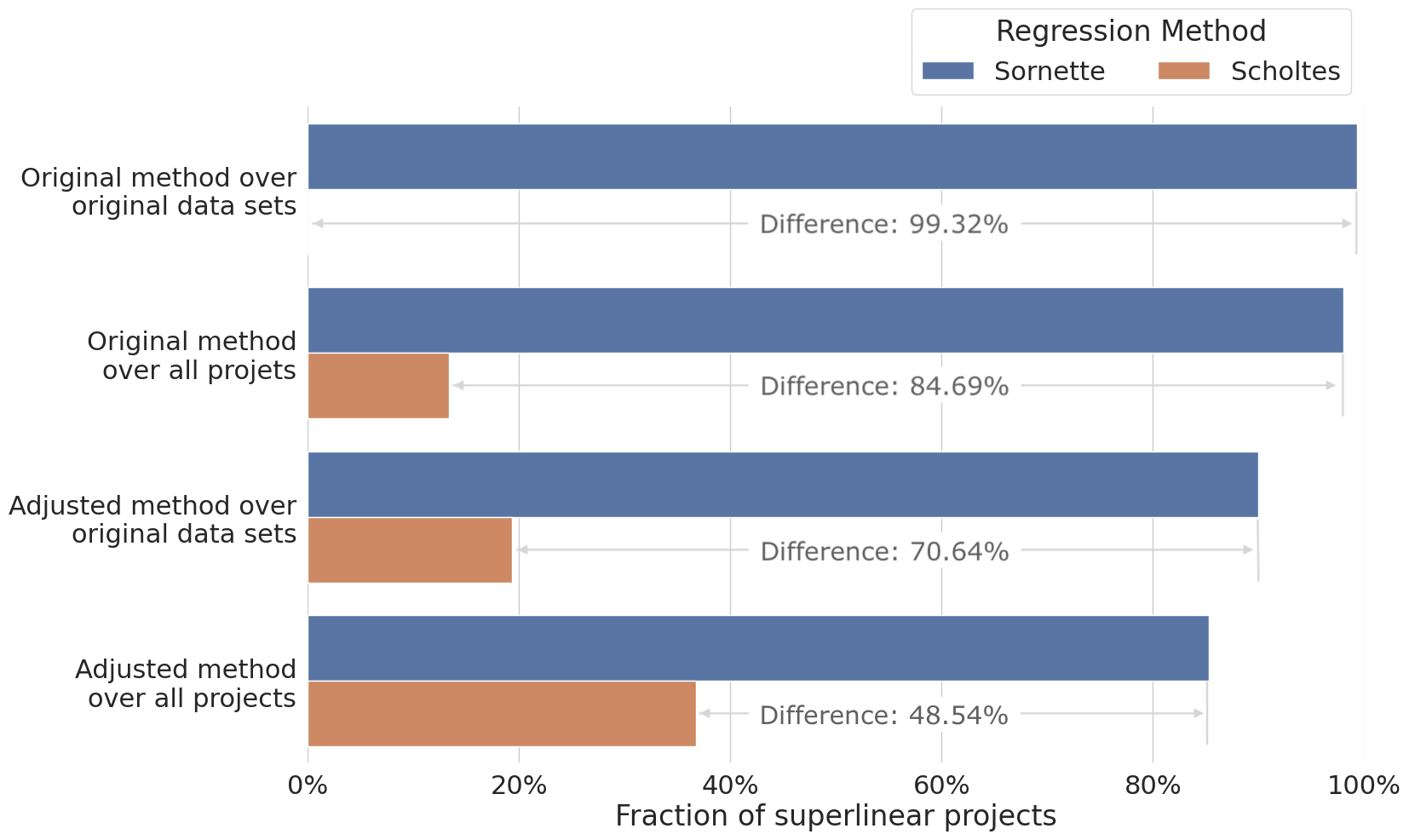}
\Description{Bar chart showing the fraction of superlinear projects.}
\caption{Summary on how selection and instrumentation biases influence the fraction of projects with a superlinear relationship for each regression method}
\label{fig:summary_superlinearity}
\end{figure}

Figure~\ref{fig:summary_superlinearity} summarizes the influences of selection and instrumentation biases. Since the population size for each analysis vary, all results are reported in percent to have a relative comparison. For example, using the original method on the original data set identified 0\%~superlinearity for the 58~projects from Scholtes. In the case of Sornette et al., the original method yielded results for 148~projects, with 99.32\%~showing superlinearity. Consequently, the difference is 99.32\%. Applying both regression methods to all projects reveals that selection bias alone does not fully account for the discrepancies, leaving a difference of 84.69\%. A sole adjustment of the regression methods approximates results better, resulting in a difference of 70.64\%. And, using the adjusted methods on all projects leaves a gap of 48.54\%. Thus, the conclusion can be drawn that selection bias as well as instrumentation biases contributed to the different conclusions drawn by Sornette et al. and Scholtes et al.

\section{Final Considerations}
\label{sec:discussion}

\subsection{Discussion}

The results of this research not only add to the Aristotle vs. Ringelmann debate, but also offer general insights into measuring productivity using Mining Software Repositories techniques. Initially, the study confirms that selection bias has an influence. Secondly, it eludes to the fact that the choice of time windows is critical, as demonstrated by the influence of the \textit{Front Load Days}. Finally, the effect of the $p$-value filter illustrated that the design of the data pipeline requires careful consideration to avoid the introduction of instrumentation biases.

This study also addresses the challenge of reproducibility in Empirical Software Engineering. In the related field of effort estimation for software development, Shepperd et al. identified only 28~replication studies~\cite{shepperd_role_2018}. Based on their assessment, there are probably hundreds or thousands of publications in this area, leading to the conclusion that only a very small percentage are actually being replicated. Cockburn et al. alert that publication bias leads to practices such as project selection, $p$ hacking, and HARKing\footnote{Hypothesizing After the Results are Known}, which might put computer science research at risk~\cite{cockburn_threats_2020}. They argue in favor of better statistical practices, experiment preregistration, openness of data, and encouragement of replications. By successfully reproducing the findings of Sornette et al. and Scholtes et al., this research contributes to the existing collection of published replication studies. Furthermore, this publication made efforts to adhere to the recommendations from Cockburn et al. by including a thorough reporting $p$-values and by providing public access to all data and calculations used in this study. 

\subsection{Lessons Learned}

The work carried out in this research also provided insight into factors that are of importance when replicating studies and offered a deeper understanding of how to reduce biases.

First, taking small steps proved to be beneficial. For example, instead of jumping right to the calculation of correlation coefficients, the reproduction of intermediate steps, such as the calculation of Levenshtein distances for single commits, already raised important questions. Second, cross-checks between different data sources, i.e. reproduced data, the results from the article and published data sets, triangulated problems, and increased confidence in the replication. Third, statistical tests also revealed to be important, leading to the detection of the $p < 0.01$ filter in Sornette's method. And last but not least, reaching out to the authors of the original papers helped to clarify implementation details. 

As this study shows, avoiding bias is not a simple undertaking. Accounting for as many factors as possible, e.g., team size, definitely addresses this issue; however, this approach has its limitations if the number of available variables gets too big or if relevant variables are either unknown or difficult to measure. Another effective method to mitigate biases is to encourage reproductions, as independent parties' reviews are an established approach to guarantee quality and reliability.

\subsection{Limitations and Future Work}

This study has some limitations, which may serve as a starting point for future work.

The influence of time window selection has been addressed only partially. Apart from the \textit{Front Load Days}, there is also the question of whether the number of days for team size and team output windows influenced the results, whether the variation in the total number of years analyzed for each project was relevant, or whether data from a more recent period would have had an impact.

Calculating the correct team size might be another factor to consider. In that regard, Gote and Zing addressed the problem of author disambiguation and provided a tool for its solution~\cite{gote_gambit_2021}. However, disambiguation has been excluded from this study to maintain consistency with the original research methodologies. Another emerging and significant question revolves around the influence of commits made by bots and AI agents.

This research has not examined confounders or other influencing variables, except for team size. Lavazza et al. identified additional factors that impact productivity, such as programming language or project type~\cite{lavazza_empirical_2018}, and Gote et al. discovered the importance of foreign code edits~\cite{gote_big_2022}. So, factors related to dimensions like team structure, source code structure, or software engineering practices may provide additional insights. 

Furthermore, the existing method of measuring productivity, which divides team output by team size, can be complemented by additional factors such as quality, maintainability, or sustainability. Recent recommendations to evaluate software development productivity in a more multidimensional manner, like the SPACE metrics, already provide a suitable framework~\cite{forsgren_space_2021}. The complexity of software development may also require complementary qualitative evaluations to improve the understanding of the underlying creative process~\cite{sadowski_no_2019}\cite{sadowski_why_2019}. Such extensions could serve as an approach to producing rigorous and quantitative research, which also offers practical insights for software development practitioners.

\section*{Acknowledgments}

We would like to thank Sornette et al. and Scholtes et al. for providing a great inspiration on how to conduct rigorous quantitative analysis of productivity in the realm of Open Source software development. In particular, we thank Thomas Maillart and Ingo Scholtes for their clarifications and feedback.

We are also grateful for the invaluable insights and discussions provided by the Software Engineering Research Group at the University of São Paulo.

\section*{Artefact Availability}

All data sets, Colab notebooks, and Python code are available on Zenodo (see \href{https://zenodo.org/records/12755951}{https://zenodo.org/records/12755951}). 


\bibliographystyle{ACM-Reference-Format}
\bibliography{references}

\end{document}